\documentclass[10pt]{article}

\usepackage[utf8]{inputenc}

\usepackage{cite}

\usepackage{nameref,hyperref}

\usepackage[right]{lineno}

\usepackage{microtype}
\DisableLigatures[f]{encoding = *, family = * }

\usepackage{changepage}

\makeatletter
\renewcommand{\@biblabel}[1]{\quad#1.}
\makeatother

\usepackage{lastpage,fancyhdr,graphicx}
\usepackage{epstopdf}
\fancyheadoffset[L]{2.25in}
\fancyfootoffset[L]{2.25in}

\usepackage{color}

\definecolor{Gray}{gray}{.25}

\usepackage{graphicx}

\usepackage{sidecap}

\usepackage{wrapfig}
\usepackage[pscoord]{eso-pic}
\usepackage[fulladjust]{marginnote}
\reversemarginpar
\usepackage{dblfloatfix}
\usepackage{wrapfig}
\usepackage{url}
\usepackage{breakurl}
\usepackage{subfig}
\usepackage{geometry}
\usepackage{multicol}
\usepackage{multirow}
\usepackage{xcolor}
\usepackage{pgfplots}
\usepackage{tikz}
\usepackage{caption}
\usepackage{tablefootnote}
\usepackage{amssymb,amsmath,amsthm}
\newtheorem{theorem}{Theorem}
\usepackage[T1]{fontenc}
\usepackage[utf8]{inputenc}
\usepackage{authblk}

\title{CryptoDL: Deep Neural Networks over Encrypted Data}

\author[1]{Ehsan Hesamifard\thanks{ehsanhesamifard@my.unt.edu}}
\author[1]{Hassan Takabi\thanks{takabi@unt.edu}}
\author[2]{Mehdi Ghasemi\thanks{mehdi.ghasemi@usask.ca}}
\affil[1]{Department of Computer Science and Engineering, 
       University of North Texas}
\affil[2]{Department of Mathematics and Statistics, University of Saskatchewan}
\begin{document}
\maketitle


\section*{Abstract}
Machine learning algorithms based on deep neural networks have achieved remarkable results and are being extensively used in different domains. However, the machine learning algorithms requires access to raw data which is often privacy sensitive. To address this issue, we develop new techniques to provide solutions for running deep neural networks over encrypted data. In this paper, we develop new techniques to adopt deep neural networks within the practical limitation of current homomorphic encryption schemes. More specifically, we focus on classification of the well-known convolutional neural networks (CNN). First, we design methods for approximation of the activation functions commonly used in CNNs (i.e. ReLU, Sigmoid, and Tanh) with low degree polynomials which is essential for efficient homomorphic encryption schemes. Then, we train convolutional neural networks with the approximation polynomials instead of original activation functions and analyze the performance of the models. Finally, we implement convolutional neural networks over encrypted data and
measure performance of the models. Our experimental results validate the soundness of our approach with several convolutional neural networks with varying number of layers and structures. When applied to the MNIST optical character recognition tasks, our approach achieves 99.52\% accuracy which significantly outperforms the state-of-the-art solutions and is very close to the accuracy of the best non-private version, 99.77\%. Also, it can make close to 164000 predictions per hour. We also applied our approach to CIFAR-10, which is much more complex compared to MNIST, and were able to achieve 91.5\% accuracy with approximation polynomials used as activation functions. These results show that CryptoDL provides efficient, accurate and scalable privacy-preserving predictions.


\section{Introduction}

Machine learning algorithms based on deep neural networks have attracted attention as a breakthrough in the advance of artificial intelligence (AI) and are the mainstream in current AI research. These techniques are achieving remarkable results and are extensively used for analyzing big data in a variety of domains such as spam detection, traffic analysis, intrusion detection, medical or genomics predictions, face recognition, and financial predictions. 
Furthermore, with increasing growth of cloud services, machine learning services can be run on cloud providers' infrastructure where training and deploying machine learning models are performed on cloud servers. Once the models are deployed, users can use these models to make predictions without having to worry about maintaining the models and the service. 
In a nutshell, this is Machine Learning as a Service (MLaaS), and several such services are currently offered including Microsoft Azure Machine Learning \cite{azureml}, Google Prediction API \cite{googlepredict}, GraphLab \cite{graphlab}, and Ersatz Labs \cite{ersatz}. However, machine learning algorithms require access to the raw data which is often privacy sensitive and can create potential security and privacy risks. 
In recent years, several studies have investigated the privacy protection of this sensitive data in different machine learning algorithms such as linear  regression \cite{mohassel}, linear  classifiers \cite{mlconfidential, bost}, decision trees \cite{statisticsmachinelearning, bost} or neural networks \cite{cryptonet2, newpaper}.


In this paper, we propose CryptoDL, a solution to run deep neural network algorithms on encrypted data and allow the parties to provide/ receive the service without having to reveal their sensitive data to the other parties. 
In particular, we focus on classification phase of deep learning algorithms.
The main components of CryptoDL are convolutional neural networks (CNNs) and homomorphic encryption (HE).
To allow accurate predictions we propose using neural networks, specifically CNNs which are extensively used in the machine learning community for a wide variety of tasks.
Recent advances in fully homomorphic encryption (FHE) enable a limited set of operations to be performed on encrypted data \cite{gentry, bgv}. This will allow us to apply deep neural network algorithms directly to encrypted data and return encrypted results without compromising security and privacy concerns. 

However, due to a number of constraints associated with these cryptographic schemes, designing practical efficient solutions to run deep neural network models on the encrypted data is challenging. 
The most notable shortcoming of HE is that operations in practical schemes are limited to addition and multiplication. Consequently, we need to adopt deep neural network algorithms within these limitations. 
The computation performed over sensitive data by deep neural network algorithms is very complex and makes it hard to support efficiently and deep neural network models cannot simply be translated to encrypted versions without modification. 
For example, in neural networks activation functions such as Rectified Linear Unit (ReLU) and Sigmoid are used as an activation function and we have to replace these functions with another function that only uses addition and multiplication such as polynomials. 

In order to have efficient and practical solutions for computations in encrypted domain, we typically need to use leveled HE schemes instead of FHE. However, a solution that builds upon these encryption schemes has to be restricted to computing low-degree polynomials in order to be practical and efficient. Approximating a function with low-degree polynomials is an important issue for running deep CNN algorithms on encrypted data when we use HE, see \cite{takabi, hesamifard}. 

\subsection{Threat Model and Problem Statement}

In this paper, we focus on the CNN, one of the most popular deep learning algorithms.
We assume that the training phase is done on the plaintext data and a model has already been built and trained.

\textbf{Problem:} \textit{Privacy-preserving Classification on Deep Convolutional Neural Networks}

The client has a previously unseen feature vector $x$ and the server has an already trained deep CNN model $w$. 
The server runs a classifier $C$ over $x$ using the model $w$ to output a prediction $C(x, w)$.
To do this, the client sends an encrypted input to the server, server performs encrypted inference, and the client gets the encrypted prediction.
The server must not learn anything about the input data or the prediction and the classification must not reveal information about the trained neural network model $w$. 

A practical solution to this problem for real-world applications should be both \textbf{accurate} (the prediction performance should be close to the prediction performance of the plaintext) and \textbf{efficient} (the running time to obtain the prediction result should be low). 

\subsection{Contributions}

In this paper, we design and evaluate a \textit{privacy-preserving classification for deep convolutional neural networks}.
Our goal is to adopt deep convolutional neural networks within practical limitations of HE while keeping accuracy as close as possible to the original model.

The most common activation functions used in CNNs are ReLU, Sigmoid, and Tanh. 
In order to achieve our goal, these functions should be replaced by HE friendly functions such as low-degree polynomials. 

The main contributions of this work are as follows.
\begin{itemize}
\item We provide theoretical foundation to prove that it is possible to find lowest degree polynomial approximation of a function within a certain error range.
\item Building upon the theoretical foundation, we investigate several methods for approximating commonly used activation functions in CNNs (i.e. ReLU, Sigmoid, and Tanh) with low-degree polynomials to find the best approximation. 
\item We utilize these polynomials in CNNs and analyze the performance of the modified algorithms. 
\item We implement the CNNs with polynomial approximations as activation functions over encrypted data and report the results for two of the widely used datasets in deep learning, MNIST and CIFAR-10.
\item Our experimental results of MNIST show that CryptoDL can achieve 99.52\% accuracy which is very close to the original model's accuracy of 99.56\%. Also, it can make close to 164000 predictions per hour. These results show that CryptoDL provides efficient, accurate, and scalable privacy-preserving predictions. 
\end{itemize}

The rest of this paper is organized as follows: In Section \ref{background1}, we provide a brief overview about the structure of HE schemes and deep CNNs. In Section \ref{polyapprox}, we describe our theoretical foundation and proposed solution for polynomial approximation in details. Section \ref{experiments} provides experimental results for CNN models over encrypted datasets followed by a discussion. In Section \ref{relatedwork}, we review related work. In Section \ref{conclusion}, we conclude the paper and discuss future work.

\section{Overview and Background Information}\label{background1}

In this section, we provide a brief introduction to HE schemes, their strengths and weaknesses which should be considered while using them for secure computation protocols. We also briefly describe deep CNNs and modifications that are required for adopting them within HE schemes. 

\subsection{Homomorphic Encryption}\label{HE}

Homomorphic encryption (HE) schemes preserve the structure of the message space such that we can perform operations such as addition and multiplication over the ciphertext space. 
Like other types of encryption schemes, an HE scheme has three main functions, $Gen$, $Enc$, and $Dec$, for key generation, encryption, and decryption, respectively. However, an HE scheme also has an evaluation function, $Eval$. Suppose we have a set of plaintext messages $\{m_{i}\}$ and relative ciphertexts $\{c_{i}\}$. Now, consider a circuit $C$. The evaluation function processes the public key $pk$, a set of ciphertexts $\{c_{i}\}$ and a circuit $C$ such that 
\begin{eqnarray*}\label{heoper}
Dec(sk,Eval(pk,C,c_{1},\cdots,c_{n})) = C(m_{1},\cdots,m_{n})
\end{eqnarray*} 
HE was first introduced in 1978 by Rivest et al. \cite{rivest}. Other researchers followed to introduce several other HE schemes \cite{paillier}. However, most of these encryption schemes have some constraints. Some of them, such as the Paillier cryptosystem \cite{paillier}, only support one operation (addition). If the encryption scheme only supports one operation, it is called Somewhat Homomorphic Encryption (SHE). 

The idea behind encryption function $Enc$ is to add a small value, called \textit{noise}, to $m$ for encrypting. Therefore, each ciphertext has a small amount of noise. When we add two ciphertexts $c_{1}$ and $c_{2}$, the result is also a ciphertext, but with noise that has grown. The $Dec$ function works correctly if this amount is less than a threshold. This threshold leads to a bound on the number of computations that can be performed over encrypted data. If an entity wants to decrease the noise, it should decrypt and encrypt the ciphertext, for decryption, it needs the secret key $sk$. For years, the community was trying to find out if there is a way to decrease the noise without having the secret key.

This question was answered in 2009 when first Fully Homomorphic Encryption (FHE) scheme was designed by Gentry \cite{gentry}.
Am FHE scheme is an HE scheme that supports circuits with arbitrary depth. In his dissertation, Gentry introduced a technique for handling an arbitrary depth of computations, called \textit{bootstrapping}.
In bootstrapping technique, the amount of noise is decreased without needing to access $sk$ \cite{gentry}. However, it has a huge computational cost and is a very slow process. This limitation makes FHE impractical for actual use. 

Recent advances in HE have led to a faster HE scheme: Leveled Homomorphic Encryption (LHE). LHE schemes do not support the bootstrapping step, so they only allow circuits with depth less than a specific threshold. If we know the number of operations before starting the computations, we can use LHE instead of FHE. The performance of LHE schemes is further improved using Single-Instruction-Multiple-Data (SIMD) techniques. Halevi et al. in \cite{helib} use this technique to create a batch of ciphertexts. So, one single ciphertext has been replaced with an array of ciphertexts in computations.

Despite the advantages of using HE schemes, they have some limitations. The first one is message space. Almost all HE schemes work with integers. Therefore, before encrypting data items, we need to convert them to integers. The second limitation is ciphertext size. The size of the message increases considerably by encryption. Another important limitation is related to the noise. After each operation, the amount of noise in ciphertext increases. Multiplication increases noise much more than addition. We should always keep the amount of noise less than the predefined threshold. The last and most important limitation is lack of division operation. In summary, only a limited number of additions and multiplications are allowed over encrypted data and therefore complex functions such as activation functions used in neural networks are not compatible with HE schemes. 

\begin{figure*}
    \centering
        \centering
        \subfloat[Convolutional Layer
        \label{cnvLayer}]{\includegraphics[width=0.4\linewidth]{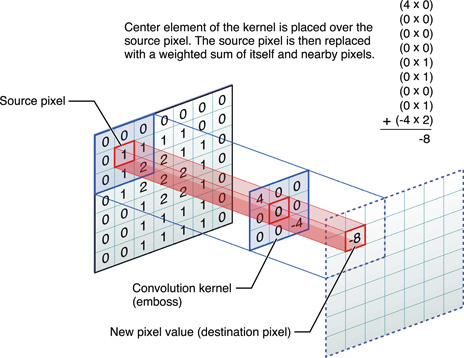}}
    \hfill
        \centering
        \subfloat[Activation Layer        
        \label{actFunc}]{\includegraphics[width=0.4\linewidth]{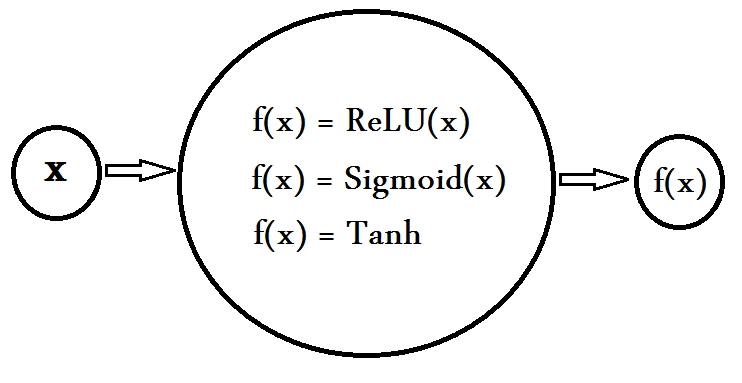}}
    \hfill
        \centering
        \subfloat[Max Pooling Layer       
        \label{maxpool}]{\includegraphics[width=0.35\linewidth]{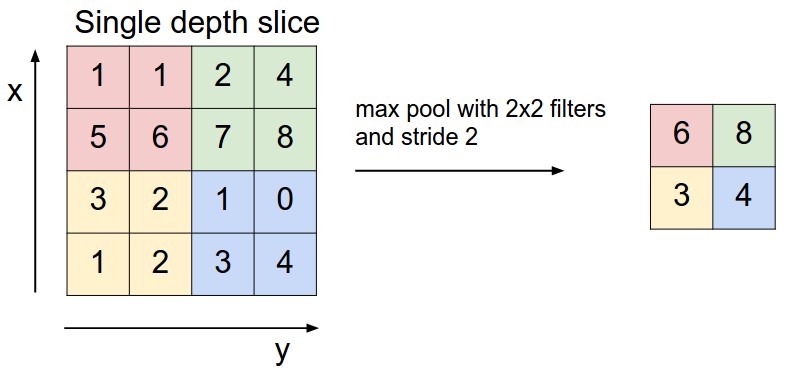}}
    \hfill
        \centering
        \subfloat[Fully Connected Layer
        \label{fullyconnected}]{\includegraphics[width=0.3\linewidth]{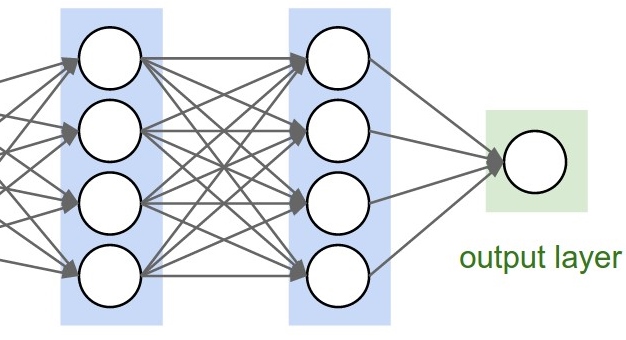}}
    \centering
    \subfloat[Dropout Layer    
    \label{dropout}]{\includegraphics[width=0.3\linewidth]{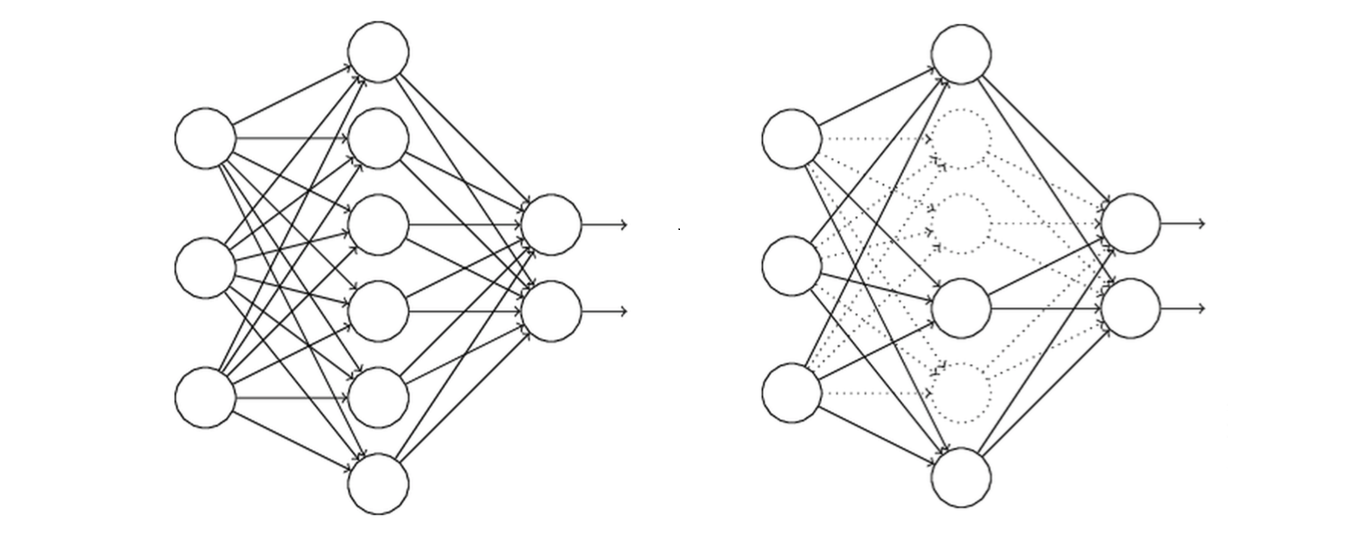}}
    \caption{Different Layers in a Convolutional Neural Network}
    \label{cnnComponents}
\end{figure*}
\setlength{\textfloatsep}{2pt}
\subsection{Deep Learning: Convolutional Neural Networks (CNNs / ConvNets)}\label{CNN}

At a high level of abstraction, a neural network is a combination of neurons arranged in ordered layers. Each neuron gets an input, operates a function on it and outputs the result of the function. The structure of this function depends on the layer to which the neuron belongs.
Besides the first layer (input layer) and the last layer (output layer), there is at least one middle layer, called hidden layer. In fully feed-forward neural networks, each neuron has a weighted connection to all neurons in the next layer. Neurons in different layers are of different types. For example, neurons in the input layer only get one input and output which is the same value. Neurons in hidden layers are more complex; they get inputs, compute the weighted summation of inputs, operate a function on the summation and then output the value of the function. These functions could be Sigmoid, Max or Mean functions and are called activation functions (or transfer functions). 

Convolutional Neural Networks (CNNs / ConvNets) are a specific type of feed-forward neural networks in which the connectivity pattern between its neurons is inspired by the organization of the animal visual cortex. They have proven to be very effective in areas such as image recognition and classification.
CNNs commonly use several distinct kinds of layers as shown in Figure \ref{cnnComponents} (adopted from http://cs231n.github.io/convolutional-networks) and described in the followings.

\subsubsection{Convolutional Layer}
The first layer in a CNN is always a convolutional layer. A convolutional layer is a set of filters that operates on the input points. For the first layer, the input is the raw image. The idea behind using convolutional layers is learning features from the data. Each filter is a $n\times n$ square (for example, $n= 3 or 5$) with a stride. We convolve the pixels in the image and calculate the dot product of the filter values and related values in the neighbor of the pixel. This step only includes addition and multiplication and we can use the same computation over the encrypted data. The stride is a pair of two numbers, for example $(2,2)$, in each step we slide the filter two units to the left or down. 

\subsubsection{Activation Layer}
After each convolutional layer, we use an activation layer which is a non-linear function. 
Every activation function takes a single number and performs a certain fixed mathematical operation on it. There are several activation functions we may encounter in practice including ReLU ($ReLU(x) = max(0,x)$), Sigmoid ($\sigma = \frac{1}{1+e^{-x}}$), and Tanh ($2\sigma(2x)-1$) functions. We cannot calculate these functions over encrypted values and we should find replacements for these functions that only include addition and multiplication operations. 

\subsubsection{Pooling Layer}
After an activation layer, a pooling layer (or sub-sampling). This layer if for sub-sampling from the data and reduces the size of data. Different kind of pooling layers are introduced in literature, two of the most popular ones are max pooling and average pooling. We cannot use max pooling because of the lack of the max operation over encrypted data. We use a scaled up version of average pooling (proposed in \cite{cryptonet2}), calculate the summation of values without dividing it by the number of values. We implement average pooling with addition only, and it does not have impact on the depth of the algorithm. 

\subsubsection{Fully Connected Layer}
Fully connected layer has the same structure like hidden layers in classic neural networks. We call this layer fully connected because each neuron in this layer is connected to all neurons in the previous layer, each connection represents by a value which called weight. The output of each neuron is the dot product of two vectors: output of neurons in the previous layers and the related weight for each neuron. 

\subsubsection{Dropout Layer}
When we train a model over training set, it's possible the final model be biased to the training set, and we get high error over the test set. This problem called over-fitting. For avoiding over-fitting during the training process, we use this specific type of layers in the CNN. In this layer, we drop out a random set of connections and set them to zero in each iteration. This dropping of values does not let the over-fitting happens in our training process and the final model is not completely fit to the training set. We need this layer only for the training step and we can remove it in the classification step. 
\subsubsection{Architecture of CNNs}
There are different ways to use the above-mentioned layers in a CNN for training a model. However, there is a common pattern for creating a CNN. The first layer is a convolutional layer, after a convolutional layer, we add an activation layer. One block of a CNN is $[Convolutional\rightarrow Activation]$ and we can use this block more than one time,  $[Convolutional\rightarrow Activation]^{n}$. After this series of block, we use an average pooling layer. This is the second block which is a combination of the first block plus an average pooling layer, $[[Convolutional\rightarrow Activation]^{n} \rightarrow Average Pooling]^{n}$. Like the first block, we can use the second block more than one times. Then, the we have one or more fully connected layers after the second block and the CNN ends with an output layer, the output of this layer is the number of classes in the dataset. This is a common pattern for creating a CNN. Figures \ref{lightCNN} and \ref{deepCNN} are two examples of CNNs with different architectures.

\section{The Proposed Privacy-preserving Classification for Deep Convolutional Neural Networks}\label{polyapprox}

Since our goal is to adopt a CNN to work within HE constraints, our focus is on operations inside the neurons. Besides activation functions inside the neurons, all other operations in a neural network are addition and multiplication, so they can be implemented over encrypted data. 
It is not possible to use activation functions within HE schemes. Hence, we should find compatible replacement functions in order to operate over encrypted data. 

The basic idea of our solution to this problem is to approximate the non-compatible functions with a compatible form so they can be implemented using HE. 
In general, most functions including activation functions used in CNNs can be approximated with polynomials which are implemented using only addition and multiplication operations. 
Hence, we aim to approximate the activation functions with polynomials and replace them with these polynomials when operating over encrypted data. 
We investigate polynomial approximation of activation functions commonly used in CNNs, namely ReLU, Sigmoid, and Tanh and choose the one that approximates each activation function the best. 

Polynomials of degree 2 have been used to substitute the Sigmoid function in neural networks \cite{cryptonet2} and polynomials of degree 3 are used to estimate the natural logarithm function \cite{securesignal}. 
However, these are specific solutions which enable us to work around certain problems, but there is no generic solution to this problem yet. 
Generally, we can approximate activation functions with polynomials from different degrees. The higher degree polynomials provide a more accurate approximation and when they replace the activation function in a CNN, lead to a better performance in the trained model. However, when operations are performed over encrypted data, a higher degree polynomial results in very slow computations. 
Therefor, a solution that builds upon HE schemes should be restricted to computing low-degree polynomials in order to be practical \cite{cryptonet1}. We need to find a trade-off between the degree of the polynomial approximation and the performance of the model.
In the following, we first provide theoretical foundation and prove that it is possible to find lowest degree polynomial approximation of a function within a certain error range. Next, we propose a solution for polynomial approximation of several activation functions (i.e. ReLU, Sigmoid, Tanh). Then, we train CNN models using these polynomials and compare the performance with the models with the original activation functions.

\subsection{Polynomial Approximation: Theoretical Foundation}

Among continuous functions, perhaps polynomials are the most well-behaved and easiest to compute. Thus, it is no surprise that mathematicians tend to approximate other functions by polynomials. Materials of this section are mainly folklore knowledge in numerical analysis and Hilbert spaces. For more details on the subject refer to \cite{math1, math2}. 

Let us denote the family of all continuous real valued functions on a non-empty compact space $X$ by $\textrm{C}(X)$.  
 Since linear combination and product of polynomials are also polynomials, we assume that $A$ is closed under addition, scalar multiplication and product and also a non-zero constant function belongs to $A$ (This actually implies that $A$ contains all constant functions). 

We say an element $f\in\textrm{C}(X)$ can be approximated by elements of $A$, if for every $\epsilon>0$, there exists $p\in A$ such that $|f(x)-p(x)|<\epsilon$ for every $x\in X$. The following classical results guarantee when every $f\in\textrm{C}(X)$ can be approximated by elements of $A$.

\begin{theorem}[Stone--Weierstrass]\label{SWT}
Every element of $\textrm{C}(X)$ can be approximated by elements of $A$ if and only if for every $x\neq y\in X$, there exists $p\in A$ such that
$p(x)\neq p(y)$.
\end{theorem}

Despite the strong and important implications of the Stone-Weierstrass theorem, it leaves computational details out and does not give a specific algorithm to generate an estimator for $f$ with elements of $A$, given an error tolerance $\epsilon$. We address this issue here.

\emph{For every $f\in\textrm{C}(X)$ and every $\epsilon>0$, there exists $p\in A$ such that $\|f-p\|_{\infty}<\epsilon$.}
Let $V$ be an $\mathbb{R}$-vector space, an \emph{inner product} on $V$ ($\langle\cdot,\cdot\rangle:V\times V\rightarrow\mathbb{R}$, see \cite{math1} for definition). The pair $(V, \langle\cdot,\cdot\rangle)$ is called an inner product space and the function $\|v\|=\langle v,v\rangle^{\frac{1}{2}}$ induces a norm on $V$. 
Every given set of linearly independent vectors can be turned into a set of orthonormal vectors (see \cite{math1} for definition) that spans the same sub vector space as the original. The Gram--Schmidt well-known theorem gives us an approach for producing such orthonormal vectors from a set of linearly independent vectors.

Now, let $\mu$ be a finite measure on $X$ and for $f,g\in\textrm{C}(X)$ define $\langle f,g\rangle=\int_Xf g d\mu$. 
This defines an inner product on the space of functions. 
Any good approximation in $\|\cdot\|_{\infty}$ gives a good $\|\cdot\|_{2,\mu}$-approximation. But generally, our interest is the other way around. 
Employing Gram--Schmidt procedure, we can find $\|\cdot\|_{2,\mu}$ within any desired accuracy, but this does not 
guarantee a good $\|\cdot\|_{\infty}$-approximation.  
In other words, \emph{``good enough $\|\cdot\|_{2,\mu}$-approximations of $f$ give good 
$\|\cdot\|_{\infty}$-approximations''}, as desired. 

Different choices of $\mu$, gives different systems of orthogonal polynomials. Two of the most popular measures are $d\mu=dx$ and $d\mu=\frac{dx}{\sqrt{1-x^2}}$. By using $d\mu=dx$ on $[-1, 1]$, the generated polynomials called Legendre polynomials and by using $d\mu=\frac{dx}{\sqrt{1-x^2}}$ on $[-1, 1]$ the generated polynomials called Chebyshev polynomials.

These two polynomial sets have different applications in approximation theory. For example, the nodes we use in polynomial interpolation are the roots of the Chebyshev polynomials and the Legendre polynomials are the coefficient of the Taylor series \cite{math1, math2}.

\subsection{Polynomial Approximation: ReLU}\label{reluApproximation}

Several methods have been proposed in the literature for polynomial approximation including Taylor series, Chebyshev polynomials, etc. \cite{math1, math2}.
We first try these approaches for finding the best approximation for the ReLU function and then present our proposed approach that provides better approximation than all these methods.
We investigate the following methods for approximating the ReLU function. 

\begin{enumerate}
    \item Numerical analysis \label{numerical}
    \item Taylor series \label{taylorm}
    \item Standard Chebyshev polynomials \label{standCheb}
    \item Modified Chebyshev polynomials \label{adopCheb}
    \item Our approach based on the derivative of ReLU function \label{simulate}
\end{enumerate}

Due to space limits, we do not provide detailed results for all these methods and only explain our conclusion for each method. 

\textbf{Method \ref{numerical}: Numerical analysis:} For this method, we generate a set of points from ReLU function and give this set of inputs to the approximation function and a constant degree for the activation function. The main shortcoming of this method is poor performance. We experimented with polynomials of degree 3 to 13 and for lower degree polynomials, the accuracy drops considerably. For achieving a good accuracy, we have to increase the degree which makes it inefficient when we are working with encrypted data. Our investigation showed that the method \ref{numerical} is not a good approach for approximating the ReLU function.

\begin{figure*}
    \centering
        \centering
        \subfloat[Approximation of ReLU using different methods
        \label{reluPic}]{\includegraphics[width=.45\linewidth]{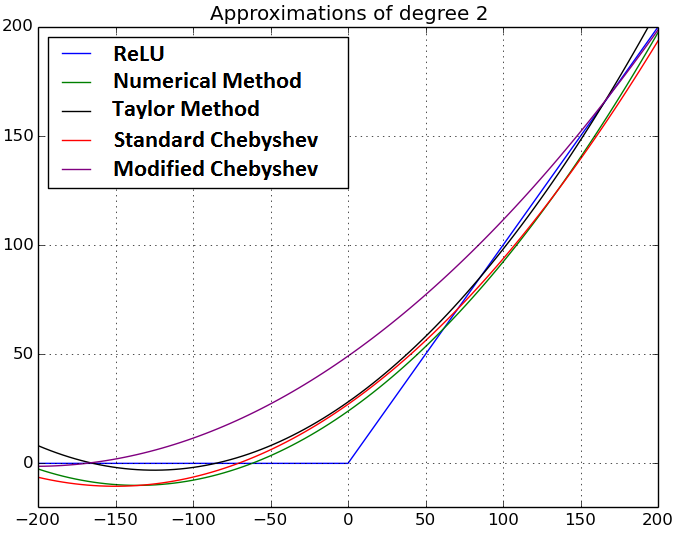}}
    \hfill
        \centering
        \subfloat[Approximation of ReLU based on our approach    
        \label{simulatePic}]{\includegraphics[width=.45\linewidth]{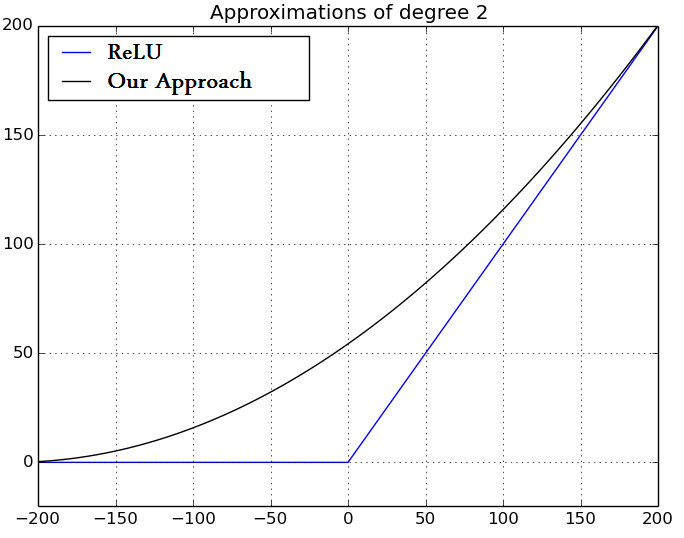}}
    \caption{Polynomial Approximation of ReLU}
    \label{relu}
\end{figure*}
\textbf{Method \ref{taylorm}: Taylor series:} In this method, we use Taylor series \cite{math1, math2}, a popular method for approximating functions. We used different degrees for approximating the ReLU function and trained the model using polynomials of different degrees. Two main issues make this method inefficient. The first issue is the high degree of polynomial approximation, although it's lower than the method \ref{numerical}, the degree is still high to be used with HE schemes. 
The second and more important issue is the interval of approximation. Basic idea of Taylor series is to approximate a function in a neighbor of a point. For the points that are not included in the input interval, the approximation error is much higher than points included in this interval. For example, in the MNIST dataset, pixel values are  integers in the interval $[0, 255]$ and this method cannot cover this interval. If we can approximate the ReLU function in a large interval, we can avoid using extra layers and this can be done using Chebyshev polynomials \cite{math1} as explained below.

\textbf{Method \ref{standCheb}: Standard Chebyshev polynomials:} Chebyshev polynomials are not as popular as the previous methods (i.e., \ref{numerical} and \ref{taylorm}). However, they have a specific feature that makes them more suitable for our problem. In this method, we approximate a function in an interval instead of a small neighborhood of a point. 
HE schemes are over integers with message space $\mathbb{Z}$; therefore, we extend the interval to be able to cover integers.  
In the standard Chebyshev polynomials, we use $d\mu=\frac{dx}{\sqrt{1-x^2}}$ as the standard norm. We approximate the ReLU function with this method and train the model based on that. As shown in Table \ref{allmethods}, the accuracy is much better than methods \ref{numerical} and \ref{taylorm}, however, it is still not a good performance in comparison with the original activation function. One way to improve the accuracy and find better approximation is to modify the measure function based on the structure of the ReLU function which is done in the next method.

\textbf{Method \ref{adopCheb}: Modified Chebyshev polynomials:} In order to simulate the structure of ReLU function, we changed the standard norm used in Chebyshev polynomials to $e^{(\frac{-1}{(1e-5+(x)^{2})})}$ and were able to achieve much better results compared to all the previous methods. The idea behind this method is that the measure for Chebyshev polynomials mainly concentrates at the end points of the interval which causes interpolation at mostly initial and end points with two singularities at both ends. While the second measure evens out through the whole real line and puts zero weight at the center. This behavior causes less oscillation in the resulting approximation and hence more similarities of derivatives with Sigmoid function. However, this improvement in performance is still not good enough. 

Let us look at why the accuracy drops significantly in the above methods.
Our goal is to approximate the ReLU function, however, note the derivative of the ReLU function is more important than the structure of the ReLU function. So, we changed our approach to the problem and focused on approximating the derivative of the ReLU function instead of approximating the ReLU function as explained below. 

\textbf{Method \ref{simulate}: Our approach based on the derivative of ReLU function:} 
All the above methods are based on simulating the activation function with polynomials. In this method, however, we use another approach and consider the derivative of the activation function because of its impact on the error calculation and updating the weights. Therefore, instead of simulating the activation function, we simulate the derivative of the activation function.
The derivative of ReLU function is like a Step function and is non-differentiable in point 0. 
If the function is continuous and infinitely derivative, we can approximate it more accurately than a non-continuous or non-infinitely differentiable function. Instead of approximating the ReLU function, we simulate the structure of derivation of the ReLU function, a Step function. Sigmoid function is a bounded, continuous and infinitely differentiable function, it also has a structure like derivative of the ReLU function in the large intervals. We approximate the Sigmoid function with the polynomial, calculate the integral of the polynomial, and use it as the activation function. 
As shown in Table \ref{allmethods}, this method achieves the best approximation of the ReLU function and we will use this method for approximation in this paper.

\begin{table}
\centering
\caption{Performance of the trained CNN using Different Approximation Methods}
\label{allmethods}
\begin{tabular}{|l|c|}
\hline
    Method & Accuracy \\
    \hline
    Numerical analysis & 56.87\% \\    
    \hline
    Taylor series & 40.28\% \\
    \hline
    Standard Chebyshev & 68.98\% \\
    \hline
    Modified Chebyshev & 88.53\%\\
    \hline
    Our Approach & 98.52\% \\
    \hline
\end{tabular}
\end{table}
\setlength{\textfloatsep}{5pt}

Figure \ref{reluPic} shows the structure of the functions generated by all the above approximations methods in comparison with the ReLU function whereas Figure \ref{simulatePic} shows only the method \ref{simulate} in comparison with the ReLU function. These two figures show that the last method simulates the structure of the ReLU function considerably better than other methods. In Figure \ref{reluPic}, for some values less than zero, the structure of the polynomial goes up and down. This behavior has impact on the performance of the model that is trained based on these polynomials. However, in Figure \ref{simulatePic}, the structure of function is almost same as the ReLU function, and for this reason, we expect to have a performance close to the ReLU function. We try to keep the degree of the polynomial as low as possible, therefore, we only work with degree 2 and degree 3 polynomials. 

\subsection{Polynomial Approximation: Sigmoid and Tanh}

In addition to the ReLU function, we also approximate two other popular activation functions: Sigmoid and Tanh. 
Approximating these two functions are more straight forward compared with the ReLU, because they are infinitely derivative. 
In this paper, we experiment with polynomial approximations of the Sigmoid function $\frac{1}{1+e^{-x}}$ and the Tanh function $Tanh(x)$, over a symmetric interval $[-l, l]$ using two different orthogonal system of polynomials. As the first choice, we consider Chebyshev polynomials on the stretched interval which come from the measure $d\mu=\frac{dx}{l\sqrt{1-(x/l)^2}}.$ 
Our second choice comes from the measure $d\mu=e^{-(l/x)^2}dx$. 

We note that the measure for Chebyshev polynomials mainly concentrates at the end points of the interval which causes interpolation at mostly initial and end points with two singularities at both ends. While the second measure evens out through the whole real line and puts zero weight at the center. This behavior causes less oscillation in the resulting approximation and hence more similarities of derivatives with Sigmoid function.

Now that we have found polynomial approximations, the next step is training CNN models using these polynomials and comparing the performance of the model with the original models.

\subsection{CNN Model 1 with Polynomial Activation Function}

In order to evaluate effectiveness of different approximation methods, we use a CNN and the MNIST dataset \cite{mnist} for our experiments. The MNIST dataset consists of 60,000 images of hand written digits. Each image is a 28x28 pixel array, where value of each pixel is a positive integer in the range $[0,255]$.
We used the training part of this dataset, consisting of 50,000 images, to train the CNN and the remaining 10,000 images for testing.

\begin{figure}
\centering
\includegraphics[width=0.8\linewidth]{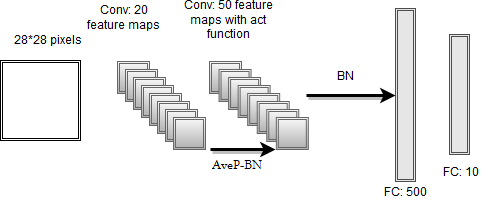}
\caption{CNN Model 1 (AveP and BN stand for Average Pooling and Batch normalization).}
\label{lightCNN}
\end{figure}

The architecture of the CNN we use is shown in Figure \ref{lightCNN}.
This CNN has similar architecture to the CNN used in \cite{cryptonet2} and the light CNN used in \cite{newpaper}. This will allow us to provide direct comparison with those work.

We train the CNN using Keras library \cite{keras} on the MNIST dataset. 
We train different models using each approximation method discussed above. We use polynomials of degree 2 to replace the ReLU function in the first four methods and polynomial of degree 3 for the last method. 
Table \ref{allmethods} shows accuracy of the trained model using each approximation method. 
As we can see, different approximation methods result in widely different accuracy values and as expected, our proposed method \ref{simulate} achieves the best accuracy among all the methods. 
In the rest of this paper, we only use this method for approximation of the ReLU function, and whenever approximation method is mentioned, it refers to the method \ref{simulate}. 

We train the model based on a CNN structure similar to the one used in \cite{newpaper} and \cite{cryptonet2}. We were able to achieve 98.52\% accuracy while their accuracy was 97.95\% .
The main difference comes from the polynomial approximation method used; the approach in \cite{newpaper} used Taylor series to approximate the ReLU function whereas we used our proposed method \ref{simulate}. These results show that our polynomial approximates the ReLU function better.

However, both results are still far from the state-of-the-art digit recognition problem (99.77\%). This is due to small size and simple architecture of the CNN used here. We use larger and more complex CNNs with more layers that are able to achieve accuracy much closer to the state-of-the-art as explained in section \ref{deepcnn}.

To ensure that the polynomial approximation is independent of the structure of the CNN and works well with different CNN structures, we change the structure of the CNN and train models based on the new one. 
The accuracy of the model based on this CNN with polynomial of degree 3 as the activation function is 98.38\% and close to the first structure. This shows that the behavior of the polynomial approximation of the ReLU function is robust against the changes in the structure. 
In addition, to analyze the relation between the degree of the polynomial and the performance of the model, we change the degree of the polynomial from 3 to 8 and calculate the accuracy of the model. As expected, the higher degree polynomials results in higher performance and we were able to achieve 99.21\% accuracy.  

\subsection{CNN Model 2 with Polynomial Activation Function} \label{deepcnn}

\begin{figure*}[!htb]
\centering
\includegraphics[width=1\linewidth]{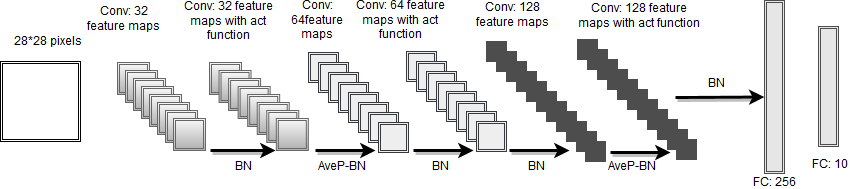}
\caption{CNN Model 2 (AveP and BN stand for Average Pooling and Batch normalization).}
\label{deepCNN}
\end{figure*}
\setlength{\textfloatsep}{5pt}
Next, we implement several CNNs with more layers and more complex structures to improve performance of the model and get closer to the state-of-the-art. We implement each model in Keras library \cite{keras} and measure the performance to find a relationship between the depth of the CNN and the accuracy of the model.

In order to provide a comparison, we use a CNN with similar structure to the one used in \cite{newpaper}. However, we start with a simpler CNN and gradually add layers to the CNN to check how the performance of the model changes as the number of layers increases and the model becomes more complex. 

First, we implement a CNN with 3 convolutional layers
The accuracy of the trained model is 99.10\% which is very close to the same CNN with the ReLU as activation function, 99.15\%. 
Next, we add an activation function to the first convolutional layer and the accuracy increases to 99.16\%. 
We then add two more convolutional layers to the CNN and train the new model. 
The accuracy of the trained model increases to 99.29\%. 
We go one step further and add an activation function after each convolutional layer. The accuracy for this CNN is 99.32\%. 
Finally, we try the same structure as the deep CNN used in \cite{newpaper}. However, the degree of our polynomial is 3 while the polynomial used in \cite{newpaper} is of degree 2. Therefore, we design our CNN, shown in Figrue \ref{deepCNN}, to have the same depth as the deep CNN used in \cite{newpaper}. \begin{table}[!htb]
\centering
\caption{Performance of the trained CNN using different Activation Functions and their Replacement Polynomials}
\label{RSTAppx}
\begin{tabular}{|c|c|c|}
\hline
     Activation Function & Original Model & Model with Polynomial \\
     \hline
     ReLU & 99.56\% & 99.52\%\\
     \hline
     Sigmoid & 98.85\% & 98.94\%\\
     \hline
     Tanh & 97.27\% & 98.15\%\\
     \hline
\end{tabular}
\end{table}

The accuracy of this model, shown in Figure \ref{deepCNN}, is 99.52\% which is very close to the accuracy of the same CNN that uses the ReLU function, 99.56\%. 
Our accuracy is higher than both methods in \cite{cryptonet2} (98.95\%) and \cite{newpaper} (99.30\%) over plaintext for a CNN with the same structure. We also train CNNs with Sigmoid and Tanh as activation functions. To provide a comparison, we use the CNN model 2 (Figure \ref{deepCNN}) and only change the activation functions to Sigmoid and Tanh instead of the ReLU function.
The results shown in Table \ref{RSTAppx}.


\section{Experimental Results: Deep Convolutional Neural Network over Encrypted Data}\label{experiments}

In this section, we present results of implementing adopted version of CNNs (ReLU is replaced with polynomial approximation) over encrypted data. 
We train the models using plaintext data and measure the accuracy of the built model for classification of encrypted data. 
We used HELib \cite{helib} for implementation and all computations were run on a computer with 16GB RAM, Intel Xeon E5-2640, 2.4GHz and Ubuntu 16.04. 




We use the CNN models trained in the previous section. We give encrypted inputs to the trained networks and measure the accuracy of the outputs. 
We implement the CNN Model 1 (Figure \ref{lightCNN}). Thanks to the SIMD feature in the HELib, in each round of classification, we can classify a batch of encrypted images. We measure the running time for encryption and sending data from the client to the server. We also measure the running time for classifying this encrypted batch as well as amount of the data transferred in the process. 

First, the encryption scheme is initiated in the client side using the the plaintext base $p$ a 16-digit prime number, $L=6$ and $k=80$ (security level which is equivalent to AES-128). Then, the input images were encrypted. As the HELib supports SIMD operations, the images were encrypted in batches. So, for each batch of images - each having the size of $28\times28$ pixels, we obtained one set of ciphertext representation of size $28\times28$. Pixel values from each location for all the images in the batch were encrypted to one ciphertext. 

Then, the encrypted images along with the encryption parameters and the public key were sent to the server side, where the server runs the CNN over the encrypted data. In our experiments, we classify a batch of ciphertext with size 8192 (the same batch size used in \cite{cryptonet2}) and provide the running time for classification. Table \ref{lightCNNtime} shows breakdown of the time it takes to apply CrytoDL to the MNIST dataset using the CNN Model 1 of Figure \ref{lightCNN}. 
We also provide time required for encryption, transferring and decryption time as shown in Table \ref{lightCNNCommtime}. As it can be seen, our approach is much faster than \cite{cryptonet2}.
 
\begin{table}
\centering
\caption{Breakdown of Running Time of CNN Model 1 (Figure \ref{lightCNN}) over Encrypted MNIST Dataset}
\label{lightCNNtime}
\begin{tabular}{|l|c|}
\hline
     Layer & Time (seconds) \\
     \hline
     Conv layer(20 feature maps) & 13.078\\
     \hline
     Average Pooling Layer & 7.630\\
     \hline
     Conv layer (50 feature maps) & 77.642\\
     \hline
     Average Pooling Layer & 6.543\\
     \hline 
     Activation layer & 9.763\\
     \hline
     2 Fully Connected (256 and 10 neurons) & 34.32\\
     \hline
\end{tabular}
\centering
\caption{Running Time for Data Transfer (seconds)}
\label{lightCNNCommtime}
\begin{tabular}{|c|c|c|}
\hline
     Layer & Our Approach (s) & CrytoNets \cite{cryptonet2} (s) \\
     \hline
     Encryption & 15.7 & 122\\
     \hline
     Communication & 320 & 570\\
     \hline
     Decryption & 1 & 5\\
     \hline
\end{tabular}
\end{table}
\setlength{\textfloatsep}{5pt}

\subsection{Comparison with the state-of-the-art Solutions} \label{comp-sota}

\begin{table*}
\begin{center}
\caption{Comparison with the state-of-the-art Solutions}
\label{comparison}
\begin{tabular}{ |c|c|c|c|c|c|c| } 
\hline
Dataset & Criteria & Our Approach & CryptoNets \cite{cryptonet2}& \cite{newpaper}$^{*}$ & DeepSecure \cite{deepsecure} & SecureML \cite{mohassel}\\
\hline
\multirow{4}{4em}{MNIST} 
& Accuracy      & 99.52\% & 98.95\% &  99.30\%$^{*}$ & 98.95\%  & 93.4\%\\ 
\cline{2-7}
& Run Time (s)     & 320    & 697    & N/A      & 10649$^{**}$ & N/A\\ 
\cline{2-7}
& Data Transfer & 336.7MB & 595.5MB   & N/A      &  722GB$^{**}$ & N/A\\ 
\cline{2-7}
& \# p/h    &  163840 & 51739 & N/A & 2769$^{**}$ & N/A\\ 
\hline
\multirow{4}{4em}{CIFAR-10} 
& Accuracy      & 91.5\%$^{***}$ & N/A & N/A & N/A & N/A\\ 
\cline{2-7}
& Run Time (s)     & 11686 & N/A & N/A & N/A & N/A\\ 
\cline{2-7}
& Data Transfer & 1803MB & N/A & N/A & N/A & N/A\\ 
\cline{2-7}
& \# p/h  &  2524  & N/A & N/A & N/A & N/A\\ 
\hline

\hline
\end{tabular}
\end{center}
\footnotesize
\begin{flushleft}
\hspace{0.2cm}*: The model is implemented over plaintext and results over encrypted data is not reported.\\
\hspace{0.2cm}**: The values are extrapolated.\\
\hspace{0.2cm}***: The model is trained over plaintext with polynomials as activation function.
\end{flushleft}
\end{table*}
\setlength{\textfloatsep}{1pt}

In this section, we compare our results with the state-of-the-art privacy-preserving classification of neural networks. This includes approaches based on the HE as well as secure multi-party computation (SMC) as shown in Table \ref{comparison}. 

The two closest work to our approach are CryptoNets \cite{cryptonet2} and \cite{newpaper}.
CryptoNets uses HE and implements a CNN with two convolutional layers and two fully connected layers. 
It assumes Sigmoid is used as the activation function and replaces the max pooling with scaled mean-pooling and the activation function with square function. As it can been seen in Table \ref{comparison}, our approach significantly outperforms CryptoNets in all aspects. Note that to provide a fair comparison, we use machines with similar configuration (Intel Xeon E5-1620 CPU running at 3.5GHz with 16GB of RAM in CryptoNets and Intel Xeon E5-2640, 2.4GHz with 16GB RAM in our case) for the experiments.
Chabanne et al. \cite{newpaper} use Taylor series for approximating the ReLU function and also use batch normalization layers for improving the performance of the model. 
They don't provide any results over the encrypted data. So, we can't provide comparison w.r.t to the performance measures (e.g., run time, amount of data transferred, number of predictions) over encrypted data but our approach provide much better accuracy. 

DeepSecure \cite{deepsecure} and SecureML \cite{mohassel} are two recent works based on SMC techniques.
Darvish et al. \cite{deepsecure} present DeepSecure that enables distributed clients (data owners) and cloud servers, jointly evaluate a deep learning network on their private assets. It uses Yao's Garbled Circuit (GC) protocol to securely perform deep learning. They perform experiments on MNIST dataset and report the results. As shown in Table \ref{comparison}, our approach significantly outperforms DeepSecure in all aspects.
Note that in \cite{deepsecure}, authors provide the communication and computation overhead for one instance, and the proposed protocol classifies one instance at each prediction round. Our approach, on the other hand, can classify a bath of instances in each round with size 8192 or larger. To provide a fair comparison, we extrapolate the running time and number of communications reported accordingly. 
Mohassel and Zhang \cite{mohassel} present SecureML that aims to develop privacy-preserving training and classification of neural networks using SMC techniques. In their proposed approach, a data owner shares the data with two servers and the servers run the machine learning algorithm using two-party computation (2PC) technique. However, they can only implement a very simple neural network with 2 hidden layers with 128 neurons in each layer, without any convolutional layers and as shown in the Table \ref{comparison}, the accuracy is very low. Implementing a CNN with their approach is not practical and hence we cannot provide comparison w.r.t to CNNs.
To provide a comparison, we implemented the same neural network (2 hidden layers with 128 neurons in each layer and without any convolutional layers) using our approach. They report 14 seconds as the running time for 100 instances, and the running time of our approach is 12 seconds. It is also worth noting that by increasing the size of the batch input, the running time increases sub-linearly in \cite{mohassel} whereas in our solution, the running time remains the same for larger sizes of the batch input. 
Additionally, unlike \cite{mohassel} our solution does not need any communications between the client and the server for providing privacy-preserving predictions. 

Generally, the SMC-based solutions have a big shortcoming which is the very large number of communications since we need interactions between client and server for each operation. For example, DeepSecure has a huge communication cost of 722GB for a relatively small network (CNN Model 1) whereas CryptoNets's communication cost is 595.5MB and ours is only 336.7MB for the same network. 
Also, since the client participates in the computations, information about the model could possibly leak. For example, the client can learn information such as the number of layers in the CNN, the structure of each layer and the activation functions. 


\subsection{CIFAR-10 Results}\label{cifar}

\begin{figure*}
        \centering
        \includegraphics[width=1\linewidth]{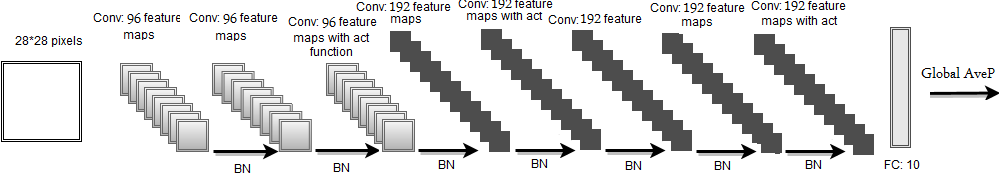}
        \caption{The architecture of CNN for CIFAR-10 classification}
        \label{cifar10}
\end{figure*}

To further show applicability of our proposed approach for more complicated network architectures, we use CIFAR-10 \cite{cifar10} which is one of the widely used benchmark dataset for deep learning, to train a CNN and implement it over encrypted data.
The CIFAR-10 dataset consists of 60000 $32\times32$ colour images categorized in 10 classes, with 6000 images per class. There are 50000 training and 10000 test images. 
We train the CNN shown in Figure \ref{cifar10} using CIFAR-10 dataset and we achieved 91.5\% accuracy with polynomials as the activation functions whereas the accuracy with the original activation function is 94.2\%. 
As shown in Table \ref{comparison}, CIFAR-10 is much slower compared to the MNIST. This was expected since both the dataset and the CNN are much more complex.

\subsection{Discussion}\label{discussion}


\textbf{Additional Datasets:} We used MNIST to be able to provide a comparison with related work which only report results on the MNIST. Additionally, we reported results on CIFAR-10 for the first time. Our solution is independent of the dataset and can be applied to other datasets such as CIFAR-100 and ImageNet. However, these datasets usually require GPU for efficient implementation and take very long time to train. This is left to our future work.

\textbf{Training CNN over encrypted data:} Although our focus in this paper is on classification phase of CNNs, it is possible to train neural networks over encrypted data. 
If we replace all the activation functions and the loss function with polynomials, back-propagation can be computed using additions and multiplications.
However, there are several challenges in doing so as stated in \cite{cryptonet2}. 
One major challenge is computational complexity. Even when we deal with plaintext, CNNs are slow to train and the deep learning community is putting a lot of effort towards increasing performance of this training process by using sophisticated hardware such as graphics processing units (GPUs) or field programmable gate arrays (FPGAs). However, adding HE to the process will make the process much slower, and therefore using leveled HE does not seem to be practical.  
Another challenging aspect in training while using HE is designing the architecture of the CNNs.
An essential part of designing efficient CNNs is the ability of data scientists to inspect the data and the trained models, and to tune the network by correcting mislabeled items, and adding features when required. This ability will not be available when encryption is used.

\section{Related Work}\label{relatedwork}

Graepel et al. use a somewhat HE scheme to train two machine learning classifiers: Linear Mean and Fisher's Linear Discriminate (FLD) \cite{mlconfidential}. They proposed division-free algorithms to adapt to limitations of HE algorithms. They focus on simple classifiers such as the linear means classifier, and do not consider more complex algorithms. Also, in their approach, the client can learn the model, and they consider a weak security model.
Bost et al. use a combination of three homomorphic systems (Quadratic Residuosity, Piallier, and BGV schemes), and garbled circuits to provide privacy-preserving classification for three different machine learning algorithms, namely Hyperplane Decision, Naive Bayes, and Decision trees \cite{bost}. Their approach is based on SMC, considers only classical machine learning algorithms and is only efficient for small data sets. 
Our proposed approach is based only on HE, focuses on the deep learning algorithms, and is efficient for large datasets. 

Xie et al. discuss theoretical aspects of using polynomial approximation for implementing neural network in encrypted domain \cite{cryptonet1}. Building on this work, Dowlin et al. present CryptoNets, a neural network classifier on encrypted data \cite{cryptonet2}. 
Chabanne et al. improve accuracy of CryptoNets by combining the ideas of Cryptonets' solution with the batch normalization principle \cite{newpaper}. 
These two are the closest to our work and were discussed in section \ref{comp-sota}. 

\cite{deepsecure} and \cite{mohassel} are two recent works based on SMC techniques and were discussed in section \ref{comp-sota}.
Aslett et al. propose methods for implementing statistical machine learning over encrypted data and implement extremely random forests and Naive Bayes classifiers over 20 datasets \cite{statisticsmachinelearning}. In these algorithms, the {majority} of operations are addition and multiplication and they show that performing algorithms over encrypted data without any multi-party computation or communication is practical. They also analyze HE tools for us{e} in statistical machine learning \cite{reviewstatistics}. 
Several methods have been proposed for statistical analysis over encrypted data, specifically for secure computation of a $\chi^{2}$-test on genome data \cite{genome1}. 
Shortell et al. use the Taylor expansion of $ln(x)$ to estimate the natural logarithm function by a polynomial of degree 5 \cite{securesignal}. 
Livni et al. analyzed the performance of polynomial as an activation function in neural networks \cite{livni}. However, their solution cannot be used for our purpose because they approximate the Sigmoid function on the interval [-1,1] while the message space of HE schemes is integers. 
Our methods is able to generate polynomial approximation for an arbitrary interval. 

There are also a few recent work that look at privacy issues in training phase, specifically for back-propagation algorithm \cite{zhang, bu}. 
Bu et al. propose a privacy-preserving back-propagation algorithm based on BGV encryption scheme on cloud \cite{bu}. Their proposed algorithm offloads the expensive operations to the cloud and uses BGV to protect the privacy of the data during the learning process.
Zhang et al. also propose using BGV encryption scheme to support the secure computation of the high-order bakc-propagation algorithm efficiently for deep computation model training on cloud \cite{zhang}.
In their approach, to avoid a multiplicative depth too big, after each iteration the updated weights are sent to the parties to be decrypted and re-encrypted. Thus, the communication complexity of the solution is very high. Unlike these papers, our focus is on privacy-preserving classification problem.
\section{Conclusion and Future Work}\label{conclusion}

In this paper, we developed new solutions for running deep neural networks over encrypted data. 
In order to implement the deep neural networks within limitations of the HE schemes, we introduced new techniques to approximate the activation functions with the low degree polynomials.
We then used these approximation to train several deep CNNs with the polynomial approximation as the activation function over the encrypted data and measured the accuracy of the trained models.
Our results show that polynomials, if chosen carefully, are the suitable replacements for activation functions to adopt deep neural networks within the HE schemes limitations. 
We were able to achieve 99.52\% accuracy and make close to 164000 predictions per hour when we applied our approach to the MNIST dataset.
We also reported results on CIFAR-10 (for the first time to the best of our knowledge) and were able to achieve 91.5\% accuracy.
These results show that our proposed approach provides efficient, accurate, and scalable privacy-preserving predictions and significantly outperforms the state-of-the-art solutions.

For future work, we plan to implement more complex models over GPU and also study privacy-preserving training of deep neural networks in addition to the classification. 



\end{document}